\documentclass[preprint,prd,showpacs]{revtex4} 
 
\begin{document} 
\input{epsf}

\title{Energy Flux Correlations and Moving Mirrors}

\author{ L.H. Ford}
 \email[Email: ]{ford@cosmos.phy.tufts.edu} 
 \affiliation{Institute of Cosmology  \\
Department of Physics and Astronomy\\ 
         Tufts University, Medford, MA 02155}
\author{Thomas A. Roman}
  \email[Email: ]{roman@ccsu.edu}
  \affiliation{Department of  Mathematical Sciences \\
 Central Connecticut State University \\  
New Britain, CT 06050}  

\begin{abstract}
We study the quantum stress tensor correlation function for a massless 
scalar field in a flat two-dimensional spacetime containing a moving
mirror. We construct the correlation functions for right-moving and 
left-moving fluxes for an arbitrary trajectory, and then specialize
them to the case of a mirror trajectory for which the expectation value 
of the stress tensor describes a pair of delta-function pulses, one
of negative energy and one of positive energy. The flux correlation function
describes the fluctuations around this mean stress tensor, and reveals subtle
changes in the correlations between regions where the mean flux vanishes. 
\end{abstract}

\maketitle

\baselineskip=13pt

\section{Introduction}
\label{sec:intro}

In recent years, there has been a considerable amount of interest in
quantum violations of the weak energy condition and the inequalities which
constrain such violations~\cite{F78,F91,FR95,FR97,FLAN,PF971,PFGQI,FE,Fewster}.
The Fulling-Davies model~\cite{FD76,FD77} of a moving mirror in two-dimensional 
spacetime provides a useful model for the study of such violations.
In particular, when the mirror is moving with increasing acceleration to 
the right, a flux of negative energy is emitted to the right. Similarly,
when the acceleration to the right is decreasing, a positive energy
flux is emitted in this direction. This model provides a simple 
illustration of how negative energy fluxes obey quantum inequalities, and 
of the phenomenon of quantum interest~\cite{FR99}. 
The Fulling-Davies model has been used
by several authors to study the phenomenon of black hole evaporation,
as there exists a specific trajectory which produces the same outgoing 
quantum state as does an evaporating black hole in two-dimensions.
Several authors~\cite{CW87,OP01,OP03a,OP03b}
 have used the mirror model to study correlations between
particles emitted in opposite directions, which in the black hole case 
are the correlations between particles emitted to infinity and those
absorbed by the black hole.

In this paper, our interest will be the correlations between the
energy fluxes moving along different null lines, rather than correlations
between particles. The positive and negative energy fluxes emitted by a
moving mirror are really {\it mean} fluxes given by the expectation
value of the quantum stress tensor. However, the quantum state is 
not an eigenstate of the stress tensor operator, so there should be
fluctuations around this mean value. Quantum stress tensor fluctuations
and their role in gravity theory are still not well understood, but
are an active area of investigation~\cite{F82,Kuo,CH95,CCV,PH97,MV99,WF99,
BFP00,PH00,WF01,FW03}. 
Even in the Minkowski vacuum state, there will be stress tensor fluctuations
with subtle correlations. The expectation value of the stress tensor vanishes
everywhere, yet an individual measurement in a finite region can return
a negative value. Somehow these fluctuations must be constrained so as
not to produce dramatic observable effects. Similarly, one expects that
even in situations where the expectation value of the stress tensor
violates the weak energy condition, there will be restrictions upon
how much additional negative energy can be observed in a given fluctuation.
However, this is not yet well understood.

In general, the study of the effects of quantum stress tensor fluctuations
requires one to deal with the singular short distance behavior of
the stress tensor correlation function. One approach is to discuss only 
spacetime averages. It is possible by a careful treatment to give
a meaning to the formally divergent integrals of the correlation function.
An alternative approach is to restrict our attention to the correlation
between stress tensors in disjoint regions of spacetime, which will
be adopted in the present paper. We will examine a particular mirror
trajectory in detail. This is a mirror which is initially static, abruptly
begins to accelerate, emitting a delta-function pulse of negative mean energy. 
The mirror subsequently stops accelerating, emitting a larger delta-function 
pulse of positive mean energy, and henceforth moves at constant velocity.  
The expectation value of the stress tensor operator is nonzero only along
the outgoing null rays emanating from the points where the acceleration
changes abruptly, and is zero elsewhere, including during the constant
acceleration phase. Our aim is to understand the correlations between
fluctuations in the flux, including in regions where the mean flux vanishes.
Obadia and Parentani~\cite{OP03b} have previously examined the correlations
between incoming and outgoing flux fluctuations for an eternally accelerating
mirror, where the expectation value of the stress tensor vanishes everywhere.

One question of interest is the degree of correlation or anticorrelation of the
negative and positive pulses. The time-integrated energy seen by any inertial
observer must be non-negative in any individual measurement~\cite{note}.
 This follows
from the fact that the integrated flux is described by an operator whose
spectrum is bounded below by zero. Thus not only must the integral of the
expectation value of the flux be non-negative, but in each measurement
the observer must detect at least as much positive as negative energy.
Now consider the situation where a fluctuation makes the negative pulse  
considerably more negative than its mean value. There must be a
corresponding upward fluctuation in the flux somewhere else to guarantee
a net positive integrated energy. Although we partially answer this 
question in the present paper, our main focus will be to 
show that there are nontrivial changes in the correlations across pulses. 

The outline of the paper is as follows: In Sect~\ref{sec:general}, we 
construct the various correlation functions for the right and left-moving 
fluxes in the presence of a mirror undergoing arbitrary motion.
In Sect.~\ref{sec:PLA} , the explicit forms of the correlation 
functions are constructed for our specific mirror trajectory.
Various limits of these functions are studied and interpreted. 
Our results are summarized and discussed in Sect.~\ref{sec:final}.

\section{Correlation Functions for a General Mirror Trajectory}
\label{sec:general}

In this paper, we will consider a massless scalar field in two-dimensional
spacetime, for which the stress tensor operator is
\begin{equation}
T_{\mu\nu} = \varphi_{,\mu}\,\varphi_{,\nu}\, - \frac{1}{2} g_{\mu\nu}\,
              \varphi_{,\rho} \varphi^{,\rho}\,.
\end{equation}
The energy flux to the right is given by
\begin{equation}
T_{uu} = \varphi_{,u}\,\varphi_{,u}\,,
\end{equation}
and that to the left is
\begin{equation}
T_{vv} = \varphi_{,v}\,\varphi_{,v}\,,
\end{equation}
where $u=t-x$ and $v=t+x$ are null coordinates. 

We now wish to construct the stress tensor correlation functions for this
theory. First consider the correlation between two right-moving rays
\begin{equation}
C_{RR}(u,u') = \langle T_{uu}(u)\, T_{uu}(u') \rangle
             - \langle T_{uu}(u)\rangle \,\langle T_{uu}(u') \rangle \,.
\end{equation}
Here the quantum state will be taken to be the in-vacuum state, the state
with no particles at early times. For this, or any other choice of vacuum state,
one may show using Wick's theorem that
\begin{equation}
\langle \varphi_1 \varphi_2 \varphi_3 \varphi_4 \rangle =
\langle \varphi_1 \varphi_2\rangle \,\langle \varphi_3 \varphi_4 \rangle
+ \langle \varphi_1 \varphi_3\rangle \,\langle \varphi_2 \varphi_4 \rangle
+\langle \varphi_1 \varphi_4\rangle \,\langle \varphi_2 \varphi_3 \rangle\,
\end{equation}
where $\varphi_1$, {\it etc} are field operators or their derivatives.
Let $\varphi_1 = \varphi_2 = \partial_u \varphi(u)$ and
$\varphi_3 = \varphi_4 = \partial_{u'} \varphi(u')$. Then we have
\begin{equation}
C_{RR}(u,u') = 2\left[\partial_u \partial_{u'} D(y,y') \right]^2 \,,
\end{equation}
where
\begin{equation}
 D(y,y') =  \langle \varphi(y),\varphi(y') \rangle
\end{equation}
is the two-point function for the quantized scalar field for the spacetime 
points $y=(u,v)$ and $y'=(u',v')$.

Similarly, we may construct the correlation function between two left-moving
rays as
\begin{equation}
C_{LL}(v,v') = \langle T_{vv}(v)\, T_{vv}(v') \rangle
             - \langle T_{vv}(v)\rangle \,\langle T_{vv}(v') \rangle \,,
\end{equation}
and that between a right moving ray and a left-moving ray as
\begin{equation}
C_{RL}(v,u') = \langle  T_{vv}(v)\, T_{uu}(u') \rangle
        -  \langle T_{vv}(v) \rangle \,\langle T_{uu}(u')\rangle \,.
\end{equation}
These functions may be expressed in terms of the two-point function as
\begin{equation}
C_{LL}(v,v') = 2\left[\partial_v \partial_{v'} D(y,y') \right]^2 \,,
\end{equation}
and 
\begin{equation}
C_{RL}(v,u') = 2\left[\partial_{u'} \partial_{v} D(y,y') \right]^2 \,.
\end{equation}

\begin{figure}
\begin{center}
\leavevmode\epsfysize=6cm\epsffile{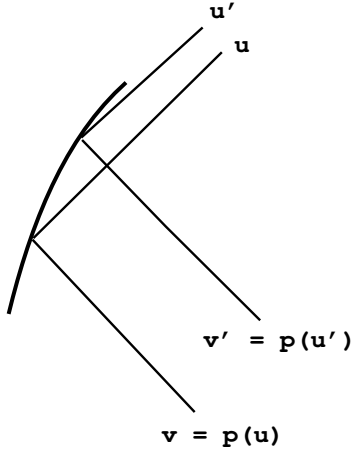}
\end{center}
\caption{The reflection of null rays by a moving mirror is illustrated.
A left-moving ray with $v = constant$ reflects from the mirror and becomes
a right-moving ray with $u = constant$, where $v = p(u)$. Another
left-moving ray $v'$ becomes a right-moving ray $u'$ with $v' = p(u')$.
Note that if the mirror is moving to the right, $\Delta v = v' - v >
\Delta u = u' - u$.}
\label{fig:u_v}
\end{figure}

The two-point function in the region to the right of a moving mirror
is given by~\cite{BD}
\begin{equation}
D = -\frac{1}{4\pi}\, \ln \left\{ 
\frac{[p(u)-p(u')-i\epsilon] (v-v'-i\epsilon)}
{[v-p(u')-i\epsilon][p(u)-v'-i\epsilon]} \right\} \,.
\end{equation}
Here the function $p(u)$ describes the reflection of null rays by the mirror:
an incoming $v = constant$ ray reflects from the mirror and becomes an
outgoing $u = constant$ ray, where the values of $u$ and $v$ are related by
$v = p(u)$, as illustrated in Fig.~\ref{fig:u_v}. Because we deal only with the 
correlations of distinct rays, for which $u \not= u'$ and $ v \not= v'$,
we can drop the $i\epsilon$ terms in the above expression.
Some explicit forms for $p(u)$ will be given in the next section.
We can now show that
\begin{equation}
C_{RR}(u,u') = \frac{[p'(u')\,p'(u)]^2}{8 \pi^2\,[p(u') - p(u)]^4} \, ,
                            \label{eq:CRR}
\end{equation}
\begin{equation}
C_{RL}(v,u') = \frac{[p'(u')]^2}{8 \pi^2\,[p(u') - v]^4} \, , \label{eq:CRL}
\end{equation}
and
\begin{equation}
C_{LL}(v,v') = \frac{1}{8 \pi^2\,(v'-v)^4} \,,  \label{eq:CLL}
\end{equation}
where $p'(u) = d p/d u$, etc.

Note that the correlation function $C_{LL}(v,v')$ is independent of the
details of the motion of the mirror, that is, of the form of the function
$p(u)$. This is to be expected. The correlations between the incoming fluxes
should not depend upon whether a mirror is present or not, and if so, whether it 
is moving. Thus the above form for $C_{LL}(v,v')$ is the same as that for
a static mirror or for empty Minkowski spacetime without a mirror. In
the limit of a static mirror, $v=p(u) = u$, and
\begin{equation}
C_{RR}(u,u') = C_{vac}(u,u') = \frac{1}{8 \pi^2\,(u'-u)^4} \,,
\end{equation}
and 
\begin{equation}
C_{RL}(v,u') = C_{static}(v,u') = \frac{1}{8 \pi^2\,(u'-v)^4} \,. 
                                     \label{eq:crl_static}
\end{equation}
The above limit of $C_{RR}(u,u')$ is the same for a static mirror and for 
empty Minkowski spacetime, but the limit for $C_{RL}(v,u')$ applies only
to a static mirror. In empty Minkowski spacetime, $C_{RL}(v,u')=0$, as 
there can only be correlations between right and left moving fluxes if there 
is a mirror present. 

We can understand the general structure of $C_{RL}(v,u')$ and $C_{RR}(u,u')$
as follows. To go from Eq.~(\ref{eq:CLL}) for $C_{LL}(v,v')$ to
Eq.~(\ref{eq:CRL}) for $C_{RL}(v,u')$, we replace $v' \rightarrow p(u')$   
and multiply by $[p'(u')]^2$. The $v' \rightarrow p(u')$ replacement simply 
describes the mapping of the left-moving to a right-moving ray as it reflects
from the mirror. The $[p'(u')]^2$ factor is a Doppler shift factor, which
describes the redshift or blueshift of an energy flux as it reflects from a 
moving mirror. Similarly, we can go from Eq.~(\ref{eq:CLL}) for $C_{LL}(v,v')$ 
to Eq.~(\ref{eq:CRR}) for $C_{RR}(u,u')$ by the double replacement
$v \rightarrow p(u)$ and $v' \rightarrow p(u')$, corresponding to the reflection 
of two left-moving rays, and multiplication by the Doppler shift factors
$[p'(u)]^2$ and $[p'(u')]^2$.

\section{A trajectory with piecewise linear acceleration} 
\label{sec:PLA} 

\subsection{Construction of $p(u)$} 
\label{sec:IIIA} 
\begin{figure}
\begin{center}
\leavevmode\epsfysize=6cm\epsffile{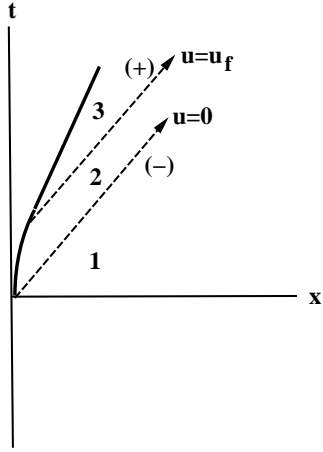}
\end{center}
\caption{A mirror starts from rest at $t=0$ and then abruptly begins to 
accelerate to the right with constant acceleration. At this time, the mirror
emits a delta-function pulse of negative energy along the line $u=0$. At
a later time, the mirror abruptly stops accelerating and emits a 
delta-function pulse of positive energy along the line $u=u_f$. Subsequently,
the mirror moves with a constant velocity. Region $1$ is $u<0$, before
the acceleration starts. Region $2$ is $0<u<u_f$, while the acceleration is
constant, but no mean energy is emitted. Region $3$ is $u>u_f$, after
the acceleration stops.  }
\label{fig:delta_pulses}
\end{figure}

As first shown by Fulling and Davies \cite{FD76,FD77}, the expectation value  
of the $uu$-component of the stress tensor to the right of the mirror 
is given by 
\begin{equation} 
\langle T_{uu} \rangle =  
\frac{1}{12 \pi}\, {(p')}^{1/2} \partial^2_u \, {(p')}^{-1/2} 
=\frac{1}{24 \pi}\, \biggl[\frac{3}{2} {\biggl(\frac{p''}{p'}\biggr)}^2  
-\frac{p'''}{p'}\biggr] \,. 
\label{eq:Tuu} 
\end{equation} 
We know that $\langle T_{uu} \rangle=0$ when the mirror moves either  
inertially or with constant acceleration. The trajectory we will construct  
corresponds to an initially static mirror at $x=0$, which begins to accelerate  
to the right at time $t=0$. The mirror then moves with constant acceleration  
until some later time when it stops accelerating and moves with constant velocity.  
The mirror will: (1)~emit a $\delta$-function pulse of $(-)$ energy when it 
starts  to accelerate at $t=0$, (2)~emit no radiation during the constant 
acceleration phase of the trajectory, and (3)~emit a $\delta$-function pulse 
of $(+)$ energy when it stops accelerating. This trajectory is depicted in 
Fig.~\ref{fig:delta_pulses}. 

Let us rewrite Eq.~(\ref{eq:Tuu}) in terms of the variable $q = p'$,  
and solve the resulting equation when we set $\langle T_{uu} \rangle=0$. We have  
\begin{equation} 
q q'' -\frac{3}{2} {(q')}^2 =0 \,. 
\end{equation}  
If we let  
\begin{equation} 
q(u) = \frac{1}{{(c_1 u+ c_0)}^2} \,, 
\end{equation} 
where $c_1, c_0$ are constants, then 
\begin{equation} 
p(u) = \frac{1}{a_1 u + a_0} + a_2 \,, 
\label{eq:pu1} 
\end{equation} 
where $a_1, a_0$ and $a_2$ are constants. 
This describes the most general, uniformly 
accelerating trajectory, for which $\langle T_{uu} \rangle=0$. 
In the special case  
when $c_1 \rightarrow \infty$, we get the solution  
\begin{equation} 
p(u) = a_3 u + a_4 \,, 
\label{eq:pu2} 
\end{equation} 
where $a_3$ and $a_4$ are constants, which describes an inertial mirror. 
 
Assume that 
\begin{equation} 
p(u)  = \left\{\matrix{u  \,, &  \,\, u \leq 0 \cr 
1/{(a_1 u + a_0)} + a_2  \,, 
& \, 0 \leq u \leq u_f \cr 
a_3 u + a_4 \,, &  \,\, u \geq u_f}\right. \,, 
\label{eq:ps} 
\end{equation} 
where $u \leq 0$ and $u \geq u_f$ correspond to the initial and  
final inertial phases of the trajectory, and $0 \leq u \leq u_f$  
represents the constant acceleration part of the trajectory.  
Continuity of $p$ and $p'$ at $u=0$ and $u=u_f$ yields 
\begin{eqnarray} 
a_1 &=& -a_0^2 \,, \\ 
a_2 &=& -1/a_0 \,, \\ 
a_3 &=& 1/{(1-a_0 u_f)}^2 \,, \\ 
a_4 &=& -a_0 u_f^2/{(1-a_0 u_f)}^2 \,. 
\end{eqnarray} 
Therefore, Eq.~(\ref{eq:ps}) becomes  
\begin{equation} 
p(u)  = \left\{\matrix{u  \,, &  \,\, u \leq 0 \,\,\,\, {\rm (region \,\, 1)}\cr 
u/{(1 - a_0 u)} \,, 
& \, 0 \leq u \leq u_f \,\,\,\, {\rm (region \,\, 2)} \cr 
(u-a_0  u_f^2)/{(1-a_0 u_f)}^2 \,, &  \,\, u \geq u_f \,\,\,\, {\rm (region \,\, 3)}} 
\right. \,, 
\label{eq:ps-exp} 
\end{equation} 
and we have that 
\begin{equation} 
p'(u)  = \left\{\matrix{1  \,, &  \,\, u \leq 0 \,\,\,\, \cr 
1/{(1 - a_0 u)}^2 \,, 
& \, 0 \leq u \leq u_f \,\,\,\,  \cr 
1/{(1-a_0 u_f)}^2 \,, &  \,\, u \geq u_f \,\,\,\, } 
\right. \,, 
\label{eq:p's} 
\end{equation}
\vspace{0.2cm}   
\begin{equation} 
p''(u)  = \left\{\matrix{0  \,, &  \,\, u < 0 \,\,\,\, \cr 
2 a_0/{(1 - a_0 u)}^3 \,, 
& \, 0 < u < u_f \,\,\,\,  \cr 
0 \,, &  \,\, u > u_f \,\,\,\,} 
\right. \,, 
\label{eq:p''s} 
\end{equation} 
and 
\begin{equation} 
p'''(u)  = p'''_D(u) + 2 a_0 \delta(u) -  
2 a_0 \delta(u-u_f)/{(1-a_0 u_f)}^3 \,, 
\label{eq:p'''s} 
\end{equation} 
with  
\begin{equation} 
p'''_D(u)  = \left\{\matrix{0  \,, &  \,\, u < 0 \,\,\,\, \cr 
6 a_0^2/{(1 - a_0 u)}^4 \,, 
& \, 0 < u < u_f \,\,\,\,  \cr 
0 \,, &  \,\, u > u_f \,\,\,\,} 
\right. \,. 
\label{eq:p'''Ds} 
\end{equation} 
Substitution of our expressions for $p',p''$, and $p'''$ into  
Eq.~(\ref{eq:Tuu}) gives 
\begin{equation}  
\langle T_{uu} \rangle = \frac{1}{12 \pi} \left[-a_0 \delta(u) +  
\frac{a_0}{ 1-a_0 u_f} \,\delta(u-u_f) \right] \,.  
\label{eq:Tuu-deltas} 
\end{equation} 
The constant $a_0$ is the proper acceleration. The case  
$a_0>0$ corresponds to the acceleration of the mirror to the right,  
with an initial $(-)$ energy $\delta$-function  
pulse emitted to the right followed by a similar $(+)$ pulse to the right.  
Since the velocity of the mirror must always be less than the speed of  
light, we must require that $u_f < 1/a_0$. For $a_0<0$, the  
mirror accelerates to the left, which corresponds to an initial $(+)$  
energy pulse to the right followed by a $(-)$ energy pulse to the right. In this  
case, there is no restriction on $u_f$. The factor $1/(1-a_0 u_f)$ is  
a Doppler shift factor which is greater than $1$ for $a_0>0$ and smaller  
than $1$ for $a_0<0$. This factor guarantees than an initial $(-)$ energy pulse must  
always be followed by a more than compensating $(+)$ energy pulse, an example of the 
phenomenon of quantum interest \cite{FR99}.  
 
\subsection{Construction of the $C_{RR}$'s} 
\label{sec:CRR-construct} 

Here we will  temporarily drop the $RR$ subscripts in 
favor of numerical subscripts which indicate the region 
of location of each ray in Fig.~\ref{fig:delta_pulses}.  
For example, $C_{12}(u,u')$ is the $RR$ correlation function with $u$  
in region 1 and $u'$ in region 2. 
In the case of the moving mirror, if we substitute 
Eqs.~(\ref{eq:ps-exp}) and (\ref{eq:p's}) into Eq.~(\ref{eq:CRR}),  
we obtain the following expressions: 
\begin{equation} 
C_{11}(u,u') = C_{22}(u,u') = C_{33}(u,u')
=   \frac{1}{8 \pi^2 {(u'-u)}^4} \,, 
\label{eq:C11-etc} 
\end{equation} 
and 
\begin{eqnarray} 
C_{12}(u,u') 
&=&  \frac{1}{8 \pi^2 {(u-u'-a_0 u u')}^4} \,\\
C_{23}(u,u') 
&=&  \frac{{(1-a_0 u_f)}^4} 
{8 \pi^2 {[u {(1-a_0 u_f)}^2-(1-a_0 u)(u'-a_0 u_f^2)]}^4} \,\\ 
C_{13}(u,u') 
&=&  \frac{{(1-a_0 u_f)}^4} 
{8 \pi^2 {[u {(1-a_0 u_f)}^2-u'+a_0 u_f^2]}^4}  \,. 
\label{eq:CRR-rest} 
\end{eqnarray} 
Notice that in all cases when $u,u'$ lie in the {\it same} region,  
the correlation function, $C_{RR}$, has the same form as in the vacuum.
This includes the case $C_{22}$, where both rays are in the region of 
uniform acceleration. We show in the Appendix that uniform acceleration
is the most general trajectory for which $C_{RR}(u,u')=C_{vac}(u,u')$.
This of course includes an inertial trajectory as a special case.

\subsection{Behavior of the $C_{RR}$'s in different regions} 
\label{sec:behavior} 
 
We saw in Sect.~\ref{sec:CRR-construct} 
that in all cases when $u,u'$ lie  
in the {\it same} region, the correlation function, $C_{RR}$, has  
the same form as in the vacuum. This is a consequence  
of the fact that the most general case for which  
$C_{RR}(u,u')=C_{vac}(u,u')$ is constant acceleration, with 
constant velocity as a special subcase.  
Referring to Fig.~\ref{fig:delta_pulses}, when  
rays $u,u'$ are both in the constant velocity regions 1 or 3, the  
mirror has zero acceleration. When both rays are in region 2, the mirror  
is undergoing constant acceleration. Only pairs of outgoing rays  
which are on opposite sides of one or both energy pulses can  
have $C_{RR}(u,u') \neq C_{vac}(u,u')$.

Let us now look at $C_{RR}/C_{vac}$ for the remaining cases, where   
\begin{eqnarray} 
\frac{C_{12}(u,u')}{C_{vac}} &=&  
\frac{{(u-u')}^4}{{(u-u'-a_0 u u')}^4} \label{eq:C12_num} \,,\\ 
\frac{C_{23}(u,u')}{C_{vac}}&=&  \frac{{(1-a_0 u_f)}^4 {(u-u')}^4} 
{{[u {(1-a_0 u_f)}^2-(1-a_0 u)(u'-a_0 u_f^2)]}^4} \,,\label{eq:C23_num}\\ 
\frac{C_{13}(u,u')}{C_{vac}}&=&  \frac{{(1-a_0 u_f)}^4 {(u-u')}^4} 
{{[u {(1-a_0 u_f)}^2-u'+a_0 u_f^2]}^4} \, . \label{eq:C13/Cvac} 
\label{eq:CRR_num} 
\end{eqnarray} 
 
We will consider primarily  the case $a_0>0$, i. e.,  
an initial $(-)$ energy pulse followed by a $(+)$ energy pulse. 
The case $a_0<0$ will be discussed briefly at the end of this subsection. 
Thus we see from Eq.~(\ref{eq:C12_num})  
that $C_{12}(u,u')/C_{vac} > 1$. One can interpret  
this as saying that the $(-)$ energy pulse {\it enhances} the correlations  
of a pair of outgoing rays which lie on opposite sides of the pulse. 
 
Now consider $C_{23}(u,u')/C_{vac}$. For $u'=u_f$, this function  
is equal to $1$; it then falls monotonically as a function of $u'$,  
for $u$ fixed and $u'>u_f$. To verify the latter property, proceed as follows.  
Let  
\begin{eqnarray} 
f(u') &=& {\biggl [\frac{C_{23}(u,u')}{C_{vac}} \biggr]}^{1/4} \,\\ 
&=& \frac{(1-a_0 u_f) (u'-u)} 
{[(1-a_0 u)(u'-a_0 u_f^2)-u {(1-a_0 u_f)}^2]} \,. 
\label{eq:f(u')} 
\end{eqnarray}  
We first check that the denominator in Eq.~(\ref{eq:f(u')})  
is always positive. Since  
$(1-a_0 u)>0$ (recall that $u<1/a_0$), the denominator is a  
monotone increasing function of $u'$, for fixed $u$.  
Furthermore, at $u'=u_f$ this function is equal to  
$(1-a_0 u_f)(u_f-u)>0$, which is consistent with  
$f(u')>0$. This confirms that  
we took the correct sign for the fourth root in  
Eq.~(\ref{eq:f(u')}). The derivative of $f(u')$ is  
\begin{equation} 
f'(u')=-a_0 \frac{(1-a_0 u_f){(u_f-u)}^2} 
{{[(1-a_0 u)(u'-a_0 u_f^2)-u {(1-a_0 u_f)}^2]}^2} <0 \,. 
\end{equation} 
However, $f(u_f)=1$ while $f'(u')<0$, so $f(u')$ and hence  
$C_{23}(u,u')/C_{vac}$, decreases monotonically from $1$ as  
$u'$ increases from $u_f$. The fact that $C_{23}(u,u')/C_{vac}<1$  
for $u'>u_f$ indicates that a $(+)$ energy pulse {\it suppresses}  
correlations across the pulse. 
 
\begin{figure}
\begin{center}
\leavevmode\epsfysize=6cm\epsffile{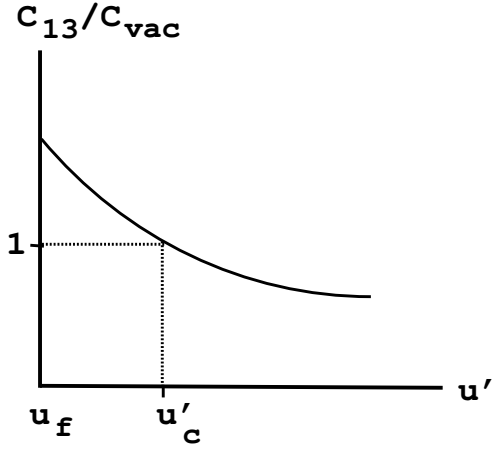}
\end{center}
\caption{The behavior of the ratio $C_{13}(u,u')/C_{vac}(u,u')$ as a function
of $u'$ for fixed $u$ is illustrated for $a_0 >0$. 
Here $u < 0$ and $u' \ge u_f$. The ratio
starts at a value greater than one at $u'=u_f$ and then falls monotonically,
passing through one at $u'=u'_c$. }
\label{fig:c13}
\end{figure}

An argument similar to that given for  
$C_{23}(u,u')/C_{vac}$ shows  
that $C_{13}(u,u')/C_{vac}$ is a monotonically  
decreasing function of $u'$, for fixed $u$. This ratio starts at a value  
greater than $1$, a fact which follows from continuity and the fact that  
$C_{12}(u,u_f)/C_{vac} > 1$. Eventually, $C_{13}(u,u'=u'_c)/C_{vac}=1$,  
at $u'_c= u_f - u(1-a_0 u_f)$, as shown in Fig.~\ref{fig:c13}.  
The fact that $C_{13}(u,u')/C_{vac}$  
eventually falls below $1$ implies that the effect of both pulses  
together is to suppress correlations, provided that the second ray $u'$  
is not too close to the $(+)$ pulse. One can think of $u'_c$ as the  
``time delay'' before the effects of the $(+)$ pulse come to dominate those of  
the $(-)$ pulse. Note that since here $u<0$, $u'_c$ grows as  
$u \rightarrow -\infty$. This behavior will be discussed in 
Sec.~\ref{sec:interp}. 
 
\subsubsection{Limits with large pulses} 
\label{sec:largepulselim} 
 
There are two ways in which the magnitudes of the pulses can  
become arbitrarily large: 1) $u_f \rightarrow 1/a_0$, where 
the $(+)$ pulse $\rightarrow \infty$ while the $(-)$ pulse 
remains fixed in magnitude;  
2) $a_0 \rightarrow \infty$, where both the $(+)$ and $(-)$ pulses  
become arbitrarily large in magnitude, but their maximum separation  
also decreases.  
 
Let us begin with the first possibility. We have that  
\begin{equation} 
C_{13}(u,u')=  \frac{{(1-a_0 u_f)}^4} 
 {8 \pi^2 {[u {(1-a_0 u_f)}^2-u'+a_0 u_f^2]}^4} \rightarrow 0 \,, 
\end{equation} 
as $u_f \rightarrow 1/a_0$, with $u,u'$, and $a_0$ fixed.  
Similarly, 
\begin{eqnarray} 
C_{23}(u,u') &=&  \frac{{(1-a_0 u_f)}^4} 
{8 \pi^2 [u {(1-a_0 u_f)}^2-(1-a_0 u)(u'-a_0 u_f^2)]^4} \,\\ 
&\sim& \frac{{(1-a_0 u_f)}^4} 
{8 \pi^2 [(1-a_0 u)(u'-a_0 u_f^2)]^4}  
\rightarrow 0 \,, 
\end{eqnarray}  
as $u_f \rightarrow 1/a_0$, if $u<u_f$, with $u,u',a_0$ fixed.  
In both of these cases, we find that the infinite $(+)$ pulse  
destroys correlations {\it across} the pulse. 
 
Now we look at the $a_0 \rightarrow \infty$ limit, 
where $u_f \rightarrow 0$. Therefore the only case 
of physical interest is $C_{13}(u,u')$. 
Let $u'> u_f = \beta/a_0$, with $\beta<1$, and $u,u'$ fixed. 
Then 
\begin{eqnarray} 
C_{13}(u,u')&=&  \frac{{(1-\beta)}^4}  
{8 \pi^2{[u {(1-\beta)}^2-u'+ \beta^2 / a_0]}^4} \,\\ 
&\sim&  
\frac{{(1-\beta)}^4}  
{8 \pi^2{[u {(1-\beta)}^2-u']}^4} \,, 
\end{eqnarray} 
as $a_0 \rightarrow \infty$.  
In this limit, the correlations do  
not vanish. Here one has large $(+)$ and $(-)$ mean pulses  
sitting nearly on top of one another, unlike in the  
$u_f \rightarrow 1/a_0$ limit, where the pulse separation  
remains finite. 
 
At this point we make some observations regarding $C_{12}(u,u_f)$.  
Let $u_f = 1/a_0 -\epsilon$. Then   
\begin{equation} 
C_{12}(u,u_f)= \frac{1} 
{8 \pi^2{[\epsilon -(1/a_0)+ \epsilon a_0 u ]}^4}  
\rightarrow \frac{a_0^4}{8 \pi^2} \,, 
\end{equation} 
as $\epsilon \rightarrow 0$. Note that in this limit,  
where the final pulse is delayed as long as possible,  
$C_{12}$ becomes almost independent of $u$. Thus we  
seem to have the peculiar situation that $C_{12}(u,u_f)$  
is almost independent of how far in the past $u$ is.  
However, we cannot {\it strictly} eliminate  
the $u$-dependence since we must always have $\epsilon >0$,  
in order to insure that the mirror's trajectory is always timelike.  
Were this not so, we might have the seemingly paradoxical  
situation in which two negative values of $u$,  
given by $(u_1,u_2)$, could be far apart so  
that $C_{11}(u_1,u_2)$ is small,  
but $C_{12}(u_1,u_f) \approx C_{12}(u_2,u_f)  
\approx a_0^4/(8 \pi^2)$.    
 
\subsubsection{The Case $a_0 <0$}
\label{sec:nega0} 

Here we will briefly discuss the case in which $a_0 <0$, 
where the mirror accelerates 
to the left, emitting an initial $(+)$ energy pulse to the right, followed later by 
a final $(-)$ energy pulse. Our previous expressions for $C_{RR}$,
Eqs.~(\ref{eq:C12_num}), (\ref{eq:C23_num}), and (\ref{eq:C13/Cvac}), are
still valid, but there is no longer any restriction on $u_f$. From
Eq.~(\ref{eq:C12_num}), we can show that
\begin{equation}
\frac{C_{12}(u,u')}{C_{vac}} \leq 0
\end{equation}
when $a_0 <0$. Thus, again the effect of a $(+)$ energy pulse is to suppress
correlations. Similarly, in this case, one finds
\begin{equation}
\frac{C_{23}(u,u')}{C_{vac}} > 0 \, ,
\end{equation}
so the correlations across the final $(-)$ energy pulse are enhanced, just as
they are for the leading $(-)$ energy pulse when $a_0 >0$. 

The behavior of $C_{13}/C_{vac}$ is somewhat more complicated, and is
essentially the inverse of the behavior for the $a_0 >0$ case depicted in 
Fig.~\ref{fig:c13}. That is, here $C_{13}/C_{vac}$ starts at $u=u_f$
with a value less than one, increases monotonically through one, and reaches
a limiting value of $(1-a_0 u_f)^4 > 1$ as $u' \rightarrow \infty$. This means
that the correlation of a ray before the leading $(+)$ energy pulse and
one long after the final $(-)$ energy pulse is enhanced, even though the
$(+)$ energy pulse has the larger magnitude.

\subsection{Construction of the $C_{RL}$'s} 
\label{sec:CRL-construct} 
Let us now consider the specific form of the $C_{RL}$ correlation  
functions for our piecewise trajectory. From Eq.~(\ref{eq:CRL}), we have 
\begin{eqnarray} 
C_{RL}(v,u' \leq 0) 
&=& \frac{1}{8 \pi^2 {(u'-v)}^4} \,\\ 
C_{RL}(v, 0 \leq u' \leq u_f)&=&  \frac{1}{8 \pi^2 {[u'-v(1-a_0 u')]}^4} \, 
\label{eq:CRL_label2}\\ 
C_{RL}(v,u' \geq u_f) 
&=&  \frac{{(1-a_0 u_f)}^4} 
{8 \pi^2 {[u' -a_0 u_f^2-v{(1-a_0 u_f)}^2]}^4}  \,. 
\label{eq:CRLs} 
\end{eqnarray} 

From Eq.~(\ref{eq:crl_static}) and the previous expressions, we obtain 
\begin{eqnarray} 
\frac{C_{RL}(v,u')}{C_{static}} &=& 1  \,   \,\,\,\,\,  
({\rm for} \,\, u' \leq 0) \,,\\ 
\frac{C_{RL}(v,u')}{C_{static}} &=&  
\frac{{(u'-v)}^4}{{[u'-v(1-a_0 u')]}^4}  \, 
 \,\,\,\, ({\rm for} \,\, 0 \leq u' \leq u_f) \label{eq:CRL/CVAC_2}   
\,,\\ 
\frac{C_{RL}(v,u')}{C_{static}}  
&=& \frac{{(1-a_0 u_f)}^4 \,{(u'-v)}^4} 
{{[u' -a_0 u_f^2-v{(1-a_0 u_f)}^2]}^4} \, \,\, 
({\rm for} \,\,\,\,\, u' \geq u_f) \,. 
\label{eq:CRL/Cvac-spec} 
\end{eqnarray} 
There are three regions for $v$:  
$v \leq 0$ - the ingoing $v$ ray arrives at the mirror  
before the acceleration phase begins;  
$0 \leq v \leq u_f/(1-a_0 u_f)$ - the ingoing ray arrives  
during the constant acceleration period;  
$v \geq u_f/(1-a_0 u_f)$ - the ingoing ray arrives after the  
acceleration has ceased.

\subsection{Interpretation of the behavior of $C_{RL}$} 
\label{sec:interp} 

In this subsection, we will discuss how the behavior of $C_{RL}$ in various
limits can be derived from kinematical considerations. We focus on the case
$a_0>0$, with just brief remarks about the case $a_0<0$.
From Eq.~(\ref{eq:CRL/CVAC_2}), we see that 
for $0 \leq u' \leq u_f < 1/a_0$ and for fixed $v<0$, with $a_0>0$, 
$C_{RL}/C_{static}$ increases as $u'$ increases. 
Let us look at this further and focus  
on $C_{RL}(v,u')$ itself. We can rewrite Eq.~(\ref{eq:CRL_label2}) as  
\begin{equation} 
C_{RL}(v,u') 
=  \frac{1}{8 \pi^2 {[u'(1+ v a_0)-v]}^4} \,. \label{eq:CRL2}
\end{equation}  
There are three possibilities: 1) $0 > v > -a_0^{-1}$,  
so $1 > 1+ v a_0 > 0$. Here $C_{RL}$ falls as $u'$ increases,  
but more slowly than does $C_{static}=1/{[8 \pi^2 {(u'-v)}^4]}$;   
2) $v=-a_0^{-1}$, which implies that $C_{RL}= const$  
independent of $u'$; and 3) $v<-a_0^{-1}$. Here $C_{RL}$  
grows as $u'$ increases. We can understand all of these behaviors  
as arising from a competition between the Doppler  
blueshift factor ${[p'(u')]}^2$, and the $1/{[p(u')-v]}^4$ factor.  
 
Recall our general expression, Eq.~(\ref{eq:CRL}), 
\begin{equation} 
C_{RL}(v,u') = \frac{{[p'(u')]}^2}{8 \pi^2 \, {[p(u')-v]}^4} \,.
\end{equation} 
As $u'$ increases, here ${[p'(u')]}^2$ increases because the mirror is  
accelerating to the right. At the same time, $p(u')$ is a monotonic increasing  
function of $u'$, so $1/{[p(u')-v]}^4 = 1/{[p(u')+ |v|]}^4$ decreases  
as $u'$ increases. This decrease simply reflects the fact that the  
incoming fluctuations are more weakly correlated as $p(u')-v$ increases.  
The two effects exactly balance when $v=-a_0^{-1}$, as seen from 
Eq.~(\ref{eq:CRL2}). We can  
understand why for more negative $v$, the effect of the  
$1/{[p(u')-v]}^4$ term is less important, as follows. The more negative  
$v$ is (larger $|v|$ is), the smaller this factor will be for a given  
value of $p(u')$. However, a given change in $p(u')$ will cause  
$1/{[p(u')-v]}^4$ to change by a smaller {\it fraction} when $|v|$ is  
larger. (Conversely, when $|v|$ is small, a small change in $p(u')$  
produces a large fractional change in $1/{[p(u')-v]}^4$.) 
 
Now let us turn to the peak value of $C_{RL}/C_{static}$ at $u'=u_f$  
for various values of $v<0$. The value at the peak is 
\begin{equation} 
\frac{C_{RL}(v,u_f)}{C_{static}(v,u_f)} =  
\frac{{[p'(u_f)]}^2 {(u_f-v)}^4} {{[p(u_f)-v]}^4}\,. 
\end{equation}  
The factor ${[p'(u_f)]}^2 = {(1-a_0 u_f)}^{-4}$ is independent of $v$,  
so we can ignore it here. From Eq.~(\ref{eq:ps-exp}) we have that  
\begin{equation} 
p(u_f)= \frac{u_f}{1-a_0 u_f} > u_f \,. 
\end{equation}
Therefore we need to look at 
\begin{equation}
\frac{{(u_f-v)}^4}{{[p(u_f)-v]}^4} = 
\frac{{(|v|+ u_f)}^4}{{[|v|+ p(u_f)]}^4} \,. 
\end{equation} 
As $|v|$ increases, this ratio varies 
smoothly from $u_f^4/{[p(u_f)]}^4 <1$ to $1$. 
This explains the increase in the peak $(u'=u_f)$ value, 
which eventually saturates at a value of ${[p'(u_f)]}^2$, which  
increases as $u_f$ increases for fixed $a_0$.  
 
Note that in the special case of $v=0$, over the interval  
$0<u'<u_f$, $C_{RL}/C_{static}=const=1$. This follows from 
Eq.~(\ref{eq:CRL/CVAC_2}). 
In this case there is no peak, since $C_{RL}/C_{static}$ 
falls monotonically for $u'>u_f$.   

Let us comment briefly on the case $a_0<0$. Here again there is a competition
between the $p(u')$ and the $p(u')-v$ factors, but each is in the opposite 
direction compared to when $a_0>0$. Thus $p(u')$ is now a redshift factor, 
but the incoming rays are more strongly correlated as $p(u')-v$ decreases.
The net effect in a typical case is to cause $C_{RL}/C_{static}$ to fall
to a minimum value, and then to rise again, the opposite behavior as
when $a_0>0$.
 
Now let us return to the case $a_0>0$ and where one ray arrives at the  
mirror before the acceleration ($v=u<0$) and one arrives there  
afterwards, $u'>u_f$. Note that here $C_{13}(u,u')\leftrightarrow 
C_{RL}(v,u')$, when $u \leftrightarrow v$, because $p'(u) = 1$ 
for $v \leq 0$. From Eq.~(\ref{eq:C13/Cvac}), we had  
\begin{equation} 
\frac{C_{13}(u,u')}{C_{vac}}= \frac{{(1-a_0 u_f)}^4 {(u-u')}^4} 
{{[u {(1-a_0 u_f)}^2-u'+a_0 u_f^2]}^4} \,. 
\label{eq:C13/Cvac_II}
\end{equation} 
This is larger than $1$ for $u'=u_f$, and then drops below $1$ at  
$u'=u'_c=u_f-u(1 -a_0 u_f)$.

\begin{figure}
\begin{center}
\leavevmode\epsfysize=6cm\epsffile{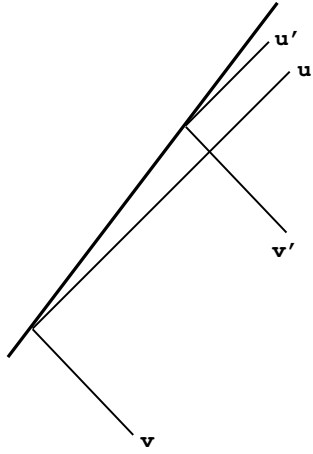}
\end{center}
\caption{The reflection of null rays by a mirror moving at a constant
velocity to the right. }
\label{fig:u_v_inertial}
\end{figure}

Consider the following two limits: 
\begin{equation} 
\frac{C_{13}(u,u')}{C_{vac}}\sim  
\left\{\matrix{1/{(1-a_0 u_f)}^4 >1   \,,  
&  \,\, u \rightarrow -\infty\,,\,\, u' \,\, {\rm fixed} \,\,\,\, \cr 
{(1 - a_0 u_f)}^4 <1 \,, 
& \, u' \rightarrow  + \infty\,,\,\, u \,\, {\rm fixed} } 
\right.  \,. 
\label{eq:C13/Cvac-lim} 
\end{equation}  
Thus for $|u|$ large, $C_{13}/C_{vac}$ starts above $1$ and  
eventually falls to a constant value $<1$ as $u'$ grows, as illustrated in
Fig.~\ref{fig:c13}.  
To understand these limits, we use the correspondence between 
$C_{13}(u,u')$ and $C_{RL}(v,u')$, and consider a mirror  
which always moves to the right at  
a constant velocity equal to the final velocity of the  
accelerated mirror. Then we have  
\begin{equation} 
p(u') = \frac{u'}{{(1-a_0 u_f)}^2} + const \,, 
\end{equation} 
and  
\begin{equation} 
C_{RL}(v,u') =  
\frac{{(1-a_0 u_f)}^{-4}}  
{{8 \pi^2 \, [u'{(1-a_0 u_f)}^{-2}-v]}^4}  
\propto  
\frac{{(1-a_0 u_f)}^4}{u'^4}\,,\,\,\,  
{\rm as} \,\, u' \rightarrow \infty \,. 
\end{equation} 
There are two competing effects here: the  
${(1-a_0 u_f)}^{-4} > 1$ factor in the numerator is  
a blueshift factor. However, it is dominated by the  
${[{(1-a_0 u_f)}^{-2}]}^4$ factor in the denominator. 
The faster the mirror is moving,  
the larger $v'-v$ is for a given $u'-u$, as illustrated in 
Fig.~\ref{fig:u_v_inertial}. This tends to cause  
the correlation function between, e. g., $v$ and $u'$,  
to be {\it smaller} because it is proportional to $1/{(v'-v)}^4$.  
This is the origin of the $1/{[p(u')-v]}^4$ factor in $C_{RL}$  
that dominates the ${[p'(u')]}^2$ factor for large $u'$.  
Another way to say this is: as we consider a sequence  
of mirrors, each moving faster to the right, a fixed  
$u',v'$ corresponds to a larger difference $v'-v$ and hence  
a {\it weaker} correlation.  
 
\begin{figure}
\begin{center}
\leavevmode\epsfysize=6cm\epsffile{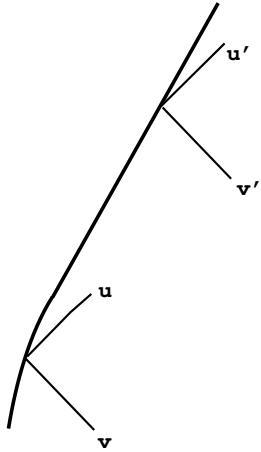}
\end{center}
\caption{Here the left-moving ray $v$ reflects from the mirror while
it is still accelerating, while the ray $v$ reflects from it long after the
acceleration has ceased. }
\label{fig:u_v_acc2}
\end{figure}

Let us now return to our piecewise accelerating trajectory.  
For the second limit in Eq.~(\ref{eq:C13/Cvac-lim}),  
as $u' \rightarrow \infty$  
for fixed $u$, the trajectory becomes very close to that  
of a mirror which is always inertial at the final velocity  
(see Fig.~\ref{fig:u_v_acc2}), and gives 
$C_{13}/C_{vac}=C_{RL}/C_{static}={(1-a_0 u_f)}^4$.  
In this case the Doppler blueshift at $u'$ is dominated by the  
stretching of the difference of the $v$ values of incoming rays.  
Similarly, the fall in the ratio after acceleration  
stops reflects the fact that it must asymptotically  
approach the value for an eternally inertial mirror at the  
final velocity. This value is smaller for faster mirrors.  

\begin{figure}
\begin{center}
\leavevmode\epsfysize=6cm\epsffile{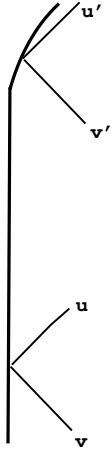}
\end{center}
\caption{Here the left-moving ray $v$ reflects from a mirror at rest. This
mirror subsequently begins to accelerate, and then the left-moving ray $v'$
reflects from it. }
\label{fig:u_v_acc}
\end{figure}

For the first limit in Eq.~(\ref{eq:C13/Cvac-lim}),  
as $u \rightarrow -\infty$ for $u'$ fixed, the trajectory  
becomes close to that of an eternally static mirror (see Fig.~\ref{fig:u_v_acc}),  
except for the blueshift factor at $u'$. The difference $v'-v$ for this  
trajectory is close to that for the eternally static mirror.  
However, there is still a ${(1-a_0 u_f)}^{-4}$ factor at $u'$, which  
produces the result for this limit.  

Which regime we are in, i.e., the first limit or the second,  
is determined by the relative magnitudes of $u{(1-a_0 u_f)}^2$  
and $u'$ in Eq.~(\ref{eq:C13/Cvac_II}). 
When $|u{(1-a_0 u_f)}^2| \gg u'$, we are in the large  
$|u|$ limit. Eventually $u'$ grows so that $|u{(1-a_0 u_f)}^2| \ll u'$,  
and we are in the other limit. However, the larger $|u|$ is, the  
longer it takes to go from one regime to the other. This explains  
why $u'_c$ grows as $|u|$ grows. Recall that this was a question which we 
posed in Sec.~\ref{sec:behavior}, as to why the decay depicted in 
Fig.~\ref{fig:c13} is slower for earlier initial rays.

\section{Summary and Discussion}
\label{sec:final}

In this paper, we have studied the flux correlation functions
$C_{RR}(u,u')$ and $C_{RL}(v,u')$ for the mirror trajectory illustrated
in Fig.~\ref{fig:delta_pulses}. In all cases, these correlation functions 
describe how the initial vacuum fluctuations are processed by the
mirror. Flux fluctuations reflect from the mirror and undergo a redshift or
blueshift. At the same time, the temporal separation between a pair
of rays can be increased or decreased by reflection. Let us recall
the main results for $C_{RR}(u,u')$, the correlation function between
a pair of right-moving fluxes which have both reflected from the mirror.
This function differs from that in the vacuum state only if $u$ and $u'$
are separated by one or both delta-function pulses in the mean stress tensor.
If they are separated by only the $(-)$ energy pulse, the correlations
are enhanced, whereas if they are separated by only the $(+)$ energy pulse, 
the correlations are suppressed compared to those in the vacuum. If both
pulses lie between $u$ and $u'$, then the correlations can be either suppressed
or enhanced, depending on the proximity of $u'$ to the second pulse. 
In the limit that the $(+)$ energy pulse becomes very large
(by letting $u_f \rightarrow 1/a_0$), the correlations across this pulse are 
destroyed and $C_{RR}(u,u') \rightarrow 0$. 

In the special case where $u$ and $u'$ lie on the mean pulses, one finds
that the correlation function takes the vacuum value, 
$C_{RR}(0,u_f)=C_{vac}(0,u_f)$. This means that the flux fluctuations around
these pulses are no more and no less correlated than are flux fluctuations
in the vacuum state. This may come as a surprise in light of the discussion in
Sect.~\ref{sec:intro} to the effect that the time integrated flux can never
fluctuate below zero. However, there are two points worth noting. The first
is that flux fluctuations in the vacuum must guarantee that the integrated
flux in any measurement is not only non-negative, but in fact zero. 
Thus the degree of correlation in the vacuum state, which suffices to 
guarantee a vanishing integrated energy in the vacuum state, also suffices 
to guarantee non-negative integrated energy flux for a moving mirror. The 
second point is that if the negative pulse fluctuates below its mean, the
compensating positive fluctuation does not have to occur near the mean 
positive pulse, but could be elsewhere.

One of the lessons of this study is that there is great deal more happening
in the accelerating mirror geometry than is revealed by the expectation
value of the stress tensor. There are subtle increases or reductions
in correlations between the flux along rays where the expectation value
vanishes.

In this paper, we have restricted our attention to the study of the correlations
of distinct regions. More generally, one would like to study the correlation
of measurements made in overlapping regions. This requires integration
of the correlation function through the singularity at coincident points. 
Integration by parts can 
be employed to define the otherwise divergent integrals~\cite{WF01,FW03}.
This will be treated in the context of moving mirror spacetimes in a future
work. 

\begin{acknowledgments}
We would like to thank Chris Fewster for helpful discussions.
  This work was supported in part by the National
Science Foundation under Grants PHY-0244898 and PHY-0139969.
\end{acknowledgments}

\appendix
\section{}

In this appendix, we will show that uniform acceleration is the most general
trajectory for which the $C_{RR}(u,u')$ correlation function has the same form
as for the vacuum state.
We want to ask when 
\begin{equation} 
C_{RR}(u,u') = \frac{{[p'(u) p'(u')]}^2}{8 \pi^2 \,{[p(u)-p(u')]}^4} = 
C_{vac}(u,u') = \frac{1}{8 \pi^2 \,{(u-u')}^4} \,, 
\label{eq:CRR=Cvac} 
\end{equation} 
which implies that 
\begin{equation} 
\frac{[p'(u) p'(u')]}{{[p(u)-p(u')]}^2} = 
\frac{1}{{(u-u')}^2} \,. \label{eq:pequ}
\label{eq:simp} 
\end{equation} 
Note that we take the $(+)$ sign when taking the square root;  
this is correct because $p(u)$ is a monotonic increasing  
function, so $p'(u)>0$ and $p'(u')>0$.  
 
If we write Eq.~(\ref{eq:simp}) as  
\begin{equation} 
-\frac{d}{du} \biggl [\frac{p'(u')}{p(u)-p(u')} \biggr ] 
= \frac{1}{{(u-u')}^2} \,, 
\end{equation} 
and integrate, we get 
\begin{equation} 
p(u)=p'(u') \,\frac{(u-u')}{[1-c_1(u')(u-u')]} + p(u') \,, 
\label{eq:p2} 
\end{equation} 
where $c_1(u')$ is a function of $u'$ alone. 
   
From this expression, we see that $p(u)$ for fixed $u'$  
must be the ratio of two linear polynomials of $u$. Let  
\begin{eqnarray} 
p(u)&=&\frac{u+b_0}{b_1 u + b_2} \,,\\ 
p'(u) &=&\frac{b_2-b_1 b_0}{{(b_1 u+ b_2)}^2} \,. 
\end{eqnarray} 
and similarly  
\begin{equation} 
p(u')=\frac{u'+ b_0}{b_1 u' + b_2} \,, \label{eq:prational}
\end{equation} 
where $b_0,b_1$, and $b_2$ are constants.  
If these relations are put back into Eq.~(\ref{eq:pequ}),  
we find that it is satisfied for all choices of  
$b_0,b_1$, and $b_2$. Thus the above $p(u)$ is the most  
general form for which $C_{RR}(u,u')=C_{vac}(u,u')$.  
From Eq.~(\ref{eq:pu1}), which can be rewritten to the form of 
Eq.~(\ref{eq:prational}), this is just the case of uniform acceleration. 
The three  
constants $b_0,b_1,b_2$ are related to the magnitude of  
the acceleration, the time at which the mirror starts  
to accelerate, and its position when it starts.


\begin{thebibliography}{28}
 
\bibitem{F78} L.H. Ford, Proc. Roy. Soc. Lond. A {\bf 364}, 227    
               (1978).    
  
\bibitem{F91} L.H. Ford, Phys. Rev. D {\bf43}, 3972 (1991).   
 
\bibitem{FR95} L.H. Ford and T.A. Roman, Phys. Rev. D {\bf 51},  
4277 (1995), gr-qc/9410043. 
  
\bibitem{FR97}  L.H. Ford and T.A. Roman, Phys. Rev. D {\bf 55},  
2082 (1997), gr-qc/9607003. 
  
\bibitem{FLAN} E.E. Flanagan,  Phys. Rev. D, {\bf 56}, 4922 (1997), 
 gr-qc/9706006. 
 
\bibitem{PF971} M.J. Pfenning and L.H. Ford, Phys. Rev. D {\bf 55},  
4813 (1997), gr-qc/9608005. 
  
\bibitem{PFGQI} M.J. Pfenning and L.H. Ford, Phys. Rev. D {\bf 57},  
                           3489 (1998), gr-qc/9710055. 
  
\bibitem{FE} C.J. Fewster and S.P. Eveson,  Phys. Rev. D {\bf 58}, 
084010 (1998), gr-qc/9805024. 
  
\bibitem{Fewster} C.J. Fewster, Class. Quantum Grav. {\bf 17}, 1897 (2000),  
gr-qc/9910060. 
 
\bibitem{FD76} S.A. Fulling and P.C.W. Davies, Proc. R. Soc. London {\bf A348},
393 (1976).

\bibitem{FD77} P.C.W. Davies and S.A. Fulling, {\it ibid}, {\bf A356},
 237 (1977).

\bibitem{FR99}  L.H. Ford and T.A. Roman, Phys. Rev. D {\bf 60}, 104018 (1999),
gr-qc/9901074. 

\bibitem{CW87} R. Carlitz and R. Willey,  Phys. Rev. D {\bf 36}, 2327 (1987).

\bibitem{OP01} N. Obadia and R. Parentani, Phys. Rev. D {\bf 64}, 044019 
(2001),  gr-qc/0103061.

\bibitem{OP03a} N. Obadia and R. Parentani, Phys. Rev. D {\bf 67}, 024021 
(2003), gr-qc/0208019.

\bibitem{OP03b} N. Obadia and R. Parentani, Phys. Rev. D {\bf 67}, 024022 
(2003), gr-qc/0209057.



\bibitem{F82}L.H. Ford, Ann. Phys (NY) \textbf{144}, 238 (1982).

\bibitem{Kuo}Chung-I Kuo and L.H. Ford, Phys. Rev. D \textbf{47}, 4510 
(1993), gr-qc/9304008.

\bibitem{CH95} E. Calzetta  and B.L. Hu, Phys. Rev. D {\bf 49}, 6636 (1993),
gr-qc/9312036; Phys. Rev. D {\bf 52}, 6770 (1995), gr-qc/9505046.

\bibitem{CCV}  E. Calzetta, A. Campos, and E. Verdaguer, Phys. Rev. D {\bf 56},
2163 (1997),  gr-qc/9704010.

\bibitem{PH97}N.G. Phillips and B.L. Hu, Phys. Rev. D \textbf{55}, 
6123 (1997), gr-qc/9611012.

\bibitem{MV99} R. Martin and E. Verdaguer, Phys. Rev. D {\bf 60}, 084008 (1999),
gr-qc/9904021.

\bibitem{WF99}C.-H. Wu and L.H. Ford, Phys. Rev. D \textbf{60}, 
104013 (1999), gr-qc/9905012.

\bibitem{BFP00}  C. Barrabes, V. Frolov, and R. Parentani, Phys. Rev. D {\bf 62},
044020 (2000),  gr-qc/0001102.

\bibitem{PH00} N.G. Phillips and B.L. Hu, Int. J. Theor. Phys. {\bf 39},
1817 (2000),  gr-qc/0004006; Phys.Rev. D {\bf 62 }, 084017 (2000),  
gr-qc/0005133. 

\bibitem{WF01} C.-H. Wu and L. H. Ford, Phys. Rev. D {\bf 64,} 045010 (2001),
quant-ph/0012144.

\bibitem{FW03} L. H. Ford and C.-H. Wu, Int. J. Theor. Phys. {\bf 42},
15 (2003), gr-qc/0102063. 

\bibitem{note} An individual measurement of the time-integrated energy
might be made, for example, by a detector switched on in the far past and
switched off in the far future.

\bibitem{BD} N.D. Birrell and P.C.W. Davies, {\it Quantum Fields in Curved
Space}, Cambridge University Press, 1982, p. 105.


\end{thebibliography}
\end{document}